\documentclass[12pt,preprint,usenatbib]{emulateapj}

\usepackage{amssymb}
\usepackage{times}
\usepackage{pifont} 

\newcommand{\teff}{$T_{\mathrm{eff}}$}

\newcommand{\numax}{$\nu_{\mathrm{max}}$}
\newcommand{\dnu}{$\Delta\nu$}

\newcommand{\kepler}{\textit{Kepler}}

\shorttitle{Amplitudes of solar-like oscillations}
\shortauthors{Stello et al.}

\begin{document}

\title{Amplitudes of solar-like oscillations:
  constraints from red giants in open clusters observed by \textit{Kepler}}  

\author{
Dennis~Stello,\altaffilmark{1} 
Daniel~Huber,\altaffilmark{1}
Thomas~Kallinger,\altaffilmark{2,3}
Sarbani~Basu,\altaffilmark{4}
Beno\^it~Mosser,\altaffilmark{5}   
Saskia~Hekker,\altaffilmark{6}
Savita~Mathur,\altaffilmark{7}
Rafael~A.~Garc{\'\i}a,\altaffilmark{8}
Timothy~R.~Bedding,\altaffilmark{1}
Hans~Kjeldsen,\altaffilmark{9} 
Ronald~L.~Gilliland,\altaffilmark{10} 
Graham~A.~Verner,\altaffilmark{11,12}
William~J.~Chaplin,\altaffilmark{11}
Othman~Benomar,\altaffilmark{1} 
S{\o}ren~Meibom,\altaffilmark{13}
Frank~Grundahl,\altaffilmark{9}
Yvonne~P.~Elsworth,\altaffilmark{11}
Joanna~Molenda-\.Zakowicz,\altaffilmark{14}
Robert~Szab\'o,\altaffilmark{15}
J{\o}rgen~Christensen-Dalsgaard,\altaffilmark{9} 
Peter~Tenenbaum,\altaffilmark{16} 
Joseph~D.~Twicken,\altaffilmark{16} 
Kamal~Uddin\altaffilmark{17} 
}
\altaffiltext{1}{Sydney Institute for Astronomy (SIfA), School of Physics, University of Sydney, NSW 2006, Australia}
\altaffiltext{2}{Department of Physics and Astronomy, University of British Columbia, 6224 Agricultural Road, Vancouver, BC V6T 1Z1, Canada}
\altaffiltext{3}{Institute for Astronomy, University of Vienna, T\"urkenschanzstrasse 17, 1180 Vienna, Austria}
\altaffiltext{4}{Department of Astronomy, Yale University, P.O. Box 208101, New Haven, CT 06520-8101}
\altaffiltext{5}{LESIA, CNRS, Universit\'e Pierre et Marie Curie, Universit\'e Denis Diderot, Observatoire de Paris, 92195 Meudon, France}
\altaffiltext{6}{Astronomical Institute ``Anton Pannekoek'', University of Amsterdam, PO Box 94249, 1090 GE Amsterdam, The Netherlands} 
\altaffiltext{7}{High Altitude Observatory, NCAR, P.O. Box 3000, Boulder, CO 80307, USA}
\altaffiltext{8}{Laboratoire AIM, CEA/DSM-CNRS, Universit\'e Paris 7 Diderot, IRFU/SAp, Centre de Saclay, 91191, Gif-sur-Yvette, France}
\altaffiltext{9}{Department of Physics and Astronomy, Aarhus University, Ny Munkegade 120, 8000 Aarhus C, Denmark}
\altaffiltext{10}{Space Telescope Science Institute, 3700 San Martin Drive, Baltimore, Maryland 21218, USA}
\altaffiltext{11}{School of Physics and Astronomy, University of Birmingham, Edgbaston, Birmingham B15 2TT, UK}
\altaffiltext{12}{Astronomy Unit, Queen Mary University of London, UK}
\altaffiltext{13}{Harvard-Smithsonian Center for Astrophysics, 60 Garden Street, Cambridge, MA, 02138, USA}
\altaffiltext{14}{Instytut Astronomiczny Uniwersytetu Wroc{\l}awskiego, ul.\ Kopernika 11,51-622 Wroc{\l}aw, Poland}
\altaffiltext{15}{Konkoly Observatory of the Hungarian Academy of Sciences, Konkoly Thege Mikl\'os \'ut 15-17, H-1121 Budapest, Hungary}
\altaffiltext{16}{SETI Institute/NASA Ames Research Center, MS 244-30, Moffat Field, CA 94035, USA}
\altaffiltext{17}{Orbital Sciences Corporation/NASA Ames Research Center, MS 244-30, Moffat Field, CA 94035, USA}

\clearpage

\begin{abstract}
Scaling relations that link asteroseismic quantities to global stellar
properties are important for gaining understanding of the intricate physics
that underpins stellar pulsation.
The common notion that all stars in an open cluster have essentially 
the same distance, age, and initial composition, implies that the stellar
parameters can be measured to much higher precision than what is usually
achievable for single stars.  This makes clusters ideal for exploring the
relation between the mode amplitude of solar-like oscillations and 
the global stellar properties.  
We have analyzed data obtained with NASA's \kepler~space telescope to
study solar-like oscillations in 100 red giant stars located in 
either of the three open clusters, NGC~6791, NGC~6819, and NGC~6811.
By fitting the measured amplitudes to predictions from simple scaling
relations that depend on luminosity, mass, and effective temperature, we find
that the data cannot be described by any power of the luminosity-to-mass
ratio as previously assumed.  As a result we provide a new improved
empirical relation which treats luminosity and mass separately.  This
relation turns out to also work remarkably well for main-sequence and
subgiant stars.
In addition, the measured amplitudes reveal the potential presence of a
number of previously 
unknown unresolved binaries in the red clump in NGC~6791 and NGC~6819,
pointing to an interesting new application for asteroseismology as a probe
into the formation history of open clusters. 
\end{abstract}

\keywords{binaries: general --- open clusters and associations: individual (NGC~6791, NGC~6819,
  NGC~6811) --- stars: fundamental parameters
  --- stars: interiors --- stars: oscillations --- techniques: photometric}

\clearpage

\section{Introduction} 
The highly complex processes involved in the excitation and damping of
stochastically excited (solar-like) oscillations 
make estimation of their amplitudes from pulsation modelling particularly
challenging \citep[e.g.~][]{Houdek06,Samadi07}.   
A scaling relation for the amplitude has therefore been of significant
interest since it was first 
introduced by \citet{KjeldsenBedding95}.  Their `$L/M$
relation', based on theoretical work by
\citet{DalsgaardFrandsen83} of near main-sequence stellar models,
suggested that the amplitude in radial velocity would simply scale as
the luminosity-to-mass ratio.  Using observations of stars made in both radial
velocity and intensity \citet{KjeldsenBedding95} also
suggested that the amplitude in intensity, $A_\lambda$, would scale as
\begin{equation}
A_\lambda=\frac{((L/L_\odot)/(M/M_\odot))^s}{\lambda/500\mathrm{nm} (T_{\mathrm{eff}}/5777\,\mathrm{K})^r}\,A_{500\mathrm{nm},\odot}, 
\label{scaling}
\end{equation}
where $s=1$, $\lambda$ is the central wavelength of
the photometric bandpass, and $A_{500\mathrm{nm},\odot}$ the observed
solar value at 500nm.
They found empirically $r=2.0$, which was a slight modification to 
$r=1.5$ derived if they assumed the stellar oscillations to be purely
adiabatic. 
Subsequent modelling by e.g. \citet{Houdek99,Samadi07} has lead to
variations of the $L/M$ relation where, in essence, different powers of the  
$L/M$ ratio have been derived ($s=0.7$--$1.3$).
Recently, \citet{Verner11} found $s=0.4$--1.0 depending on \teff~of a large
sample (642) of main-sequence and subgiant stars observed by
\kepler~\citep{Koch10}.  

The existence of solar-like oscillations in red giant stars is
now well established observationally, most recently from CoRoT
\citep[e.g.~][]{Ridder09} 
and \kepler~\citep[e.g.~][]{Gilliland10,Bedding10}, as well as
theoretically \citep{Dupret09,Montalban10,Dimauro11}.   
Despite the significantly different structures of red giants compared to
the stars and models on which the $L/M$ relation has been
founded, the absence of an alternative has also seen this relation widely
used for red giants, including several attempts to determine the best
matching exponent, $s$ \citep[][]{Stello07,Mosser10,Baudin11}.    
While the majority of results on red giants are on field stars, 
the recent clear detections in open cluster red giants emerging from
\kepler~\citep[][Paper I]{Stello10} has opened up the seismic exploration of
clusters and the advances that clusters bring to the interpretation 
of asteroseismic data (\citealt{Basu11,Hekker10a}; Miglio et al. in
prep.; Stello et al. submitted). 
In particular, stars in an open cluster are thought to share
a common distance and initial chemical composition, which allows one to
derive the stellar luminosity to much higher precision than for most field
stars.  In addition, the common age of red giant stars within each cluster
implies that they have
practically the same mass, resulting in a relatively low uncertainty on their
measured mean mass assuming there is no significant mass loss
(Miglio et al. in prep.).  Combined with high quality standard photometry
we can therefore obtain more robust predictions of the amplitudes from
scaling relations and hence investigate these in ways not possible for the
field stars observed by the current space mission CoRoT and \kepler. 

Based on only one month of \kepler~data of a single cluster,
in Paper I we already demonstrated the potential for investigating the $L/M$
relation by taking advantage of the common cluster properties of the
stellar sample.  
We now have \kepler~time-series photometry that span 10 times longer
for stars in three open clusters (NGC~6791, NGC~6819, and NGC~6811),
which exhibit distinctly different stellar masses. 
In this paper we are therefore extending considerably
the analysis of the amplitude scaling relation for solar-like oscillations.

\section{Observations, target selection \& cluster parameters}\label{observations}
The photometric time-series data were obtained between 2009 May 12 and 2010
March 20 (observing quarters 1--4), providing approximately 14,000 data
points per star obtained in the spacecraft's long-cadence mode ($\Delta t =
29.4\,$min).  A detailed description of the data reduction from raw
images to final light curves is given in
\citet{Jenkins10a,Garcia11} and Stello et al. (submitted).

Our initial star sample was the one selected by Stello et al. (submitted),
who used seismic and conventional measurements to identify
cluster membership and blending of each star.
We excluded the seismic non-members and  
further trimmed the sample by removing the brightest (largest) and
faintest stars, 
for which the measurement of the mode amplitude would not be
reliable due to: (1) difficulty in determining the noise level at low
frequency in the power spectrum of the largest stars (oscillating at very
low frequencies) and (2) low signal-to-noise and potential blending of the
faintest stars.   
The increased flux in the photometric aperture from a blending star, such as
an unresolved binary companion, will tend to reduce the relative flux
variation that we measure as the oscillation amplitude.  
To minimize this bias further, we excluded a total of 23 spectroscopic
binaries \citep{Hole09} and stars that we expected to be binaries based on
their location in the color-magnitude diagram.  
This still left a large sample of 100 stars for further analysis.
We finally investigated effects of blending of single stars based on the
results by Stello et al..  Few of the blended stars
indicated by Stello et al. showed lower than expected amplitudes, but no  
rigorous criterion for when blending had a significant impact on the
amplitude could be obtained from those results.  We therefore did not
exclude any of our remaining stars that were listed as blends. 
\begin{figure}
\includegraphics{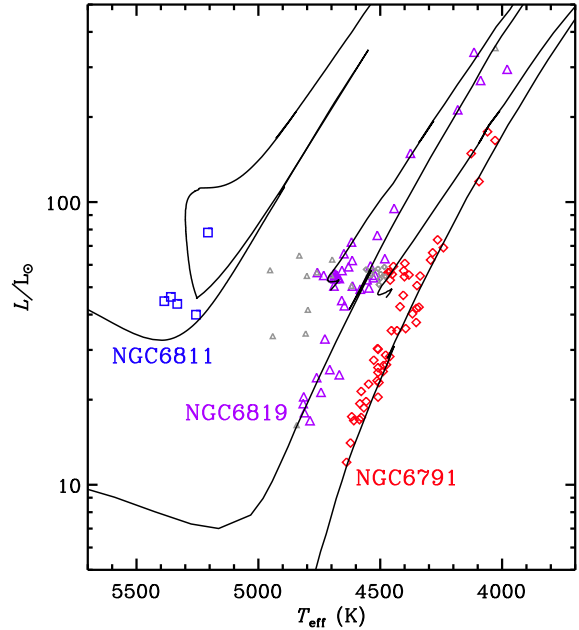}
\caption{
  H-R diagram of the selected cluster stars. Small gray symbols
  mark the known and potential binaries.  Representative isochrones from
  \citet{Marigo08} (NGC~6791: 5.6 Gyr, $Z=0.030$ and NGC~6819: 2.4 Gyr,
  $Z=0.019$) and \citet{Pietrinferni04} (NGC~6811: 500 Myr, $Z=0.008$) are
  shown to guide the eye. 
\label{cmd}} 
\end{figure} 

We adopted the luminosities, masses and effective temperatures from
Stello et al. (submitted).  We refer to \citet{Basu11} and
\citet{Hekker10a} for further details on the derivation on the mass and
effective temperature, respectively. 
In summary, the average mass (here adopted for each star) is
$1.20\pm0.01\,$M$_\odot$ (NGC~6791), $1.68\pm0.03\,$M$_\odot$ (NGC~6819),
and $2.35\pm0.04\,$M$_\odot$ 
(NGC~6811), while the luminosities and temperatures of our final sample are shown in
Figure~\ref{cmd} and have typical uncertainties of $\sim10\%$ and
$\sim2\%$, respectively.

\section{Measurement of oscillation amplitudes and \numax}\label{extract}
\begin{figure}
\includegraphics{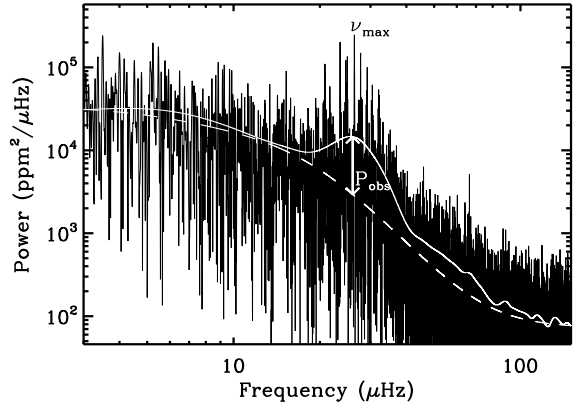}
\caption{
  Power spectrum of a typical star. The smoothed spectrum (solid white
  line) and fit to the stellar granulation background (dashed line) are shown.
  The oscillation power, $P_\mathrm{obs}$, is evaluated at the frequency of
  maximum power, \numax.   
\label{spectrum}} 
\end{figure}

Oscillation amplitudes were extracted 
by five different teams using pipelines described in 
\citet{Hekker10,Huber09,Kallinger10a,Mathur09a,MosserAppourchaux09}. 
These methods are all based on the measurement of the integrated oscillation
power, which we converted to an amplitude per radial mode.
The integrated power was found either by smoothing the power spectrum as
described by \citet{Kjeldsen07} or by fitting a Gaussian function to the
oscillation power envelope.  Figure~\ref{spectrum} shows the former. 
To obtain the amplitude per radial mode the oscillation power ($P_\mathrm{obs}$,
Figure~\ref{spectrum}) is 
multiplied by \dnu~to obtain the power per radial
order, where \dnu~is the frequency separation between consecutive radial
orders.  Finally we divided by the factor, $c$, which is the 
effective number of modes per \dnu~ 
\citep{Kjeldsen07}.  We adopted the solar value $c=3.04$ from
\citet{Bedding10a}, which agrees well with the measured mean value for red
giants (Mosser et al. submitted). 
We note that our final results (Sect.~\ref{new}) were 
not affected significantly if we adopted the recent factor by \citet{Ballot11}. 
Hence, the observed amplitude per radial mode,
$A_\mathrm{obs} (l=0)$ was derived as:
\begin{equation}
A_\mathrm{obs} (l=0)=(P_\mathrm{obs}\Delta\nu/3.04)^{1/2}.
\end{equation}
For this we normalized the power spectra according
to the amplitude-scaled version of Parseval's theorem
\citep{KjeldsenFrandsen92}, in which a sine wave of amplitude,
$A$, provides a peak in the power spectrum of $A^2$. 
The typical uncertainty in the measured amplitude is $\sim10\%$.

To explore whether the applied solar conversion factor, $c$, provided
reasonable amplitudes for red giants, we ran simulations that as input took  
pulsation frequencies derived using the ADIPLS code
\citep{DalsgaardAdipls08} for a representative set of ASTEC models
\citep{DalsgaardAstec08}.
Details of the simulator can be found in \citet{Chaplin08}.
Following \citet{Dalsgaard04}, the input mode amplitudes were scaled
relative to the radial modes using the mode inertia, $I$, as $A\propto
I^{-2}$.  
Despite significant differences in the frequency spectra of red giants
compared to the Sun, in particular the presence of many mixed modes
\citep{Dupret09,Beck11,Bedding11}, the pipelines 
returned amplitudes within 10\% of the input values.  
We regard this as acceptable given the uncertainty from intrinsic scatter
of the oscillations and the slightly different approaches for extracting
the amplitudes in each pipeline, in particular the fitting and subtraction
of the stellar granulation background (Mathur et al. submitted).     
Based on a representative set of stars, we found good agreement between
the different pipelines. 
In this paper we show the results from 
the SYD pipeline \citep{Huber09}, which provided amplitudes for the widest
range of stars, and we compare our final result with the CAN pipeline
\citep{Kallinger10a}, which exhibited the largest overlap in stellar sample
with the SYD pipeline.  Both pipelines show robust performances in their
estimation of the stellar granulation background (Mathur et al. submitted).
We refer to \citet{Verner11} and Mosser et al. (submitted) for detailed
amplitude comparisons. 

In addition to amplitude, the pipelines also measured the frequency of
maximum power, \numax~(Figure~\ref{spectrum}). The uncertainties in \numax~
are typically 1--2\%.

\section{Results}
\subsection{$A_\mathrm{obs}$ versus \numax}
As noted by \citet{Stello07,Mosser10,Huber10}, it can be convenient 
to plot the measured amplitude as a function of \numax, since the
currently adopted scaling relations predict a simple relation
between the two. 
In particular, by dividing
$A_\lambda\propto(L/M)^sT_{\mathrm{eff}}^{-r}$ (Equation~\ref{scaling}) by
(\numax$)^s\propto(M/L)^sT_{\mathrm{eff}}^{3.5s}$ \citep{Brown91}
and rearranging, we obtain  
$A_\lambda\propto\nu_{\mathrm{max}}^{-s}T_{\mathrm{eff}}^{3.5s-r}$. 
Hence, such a purely empirical plot allows one to make some inference on how
the amplitude depends on the stellar parameters $L$, $M$, and \teff~even
when those are not very well known (e.g. \citealt{Mosser10,Huber10};
Huber et al. submitted; Mosser et al. submitted).

\begin{figure}
\includegraphics{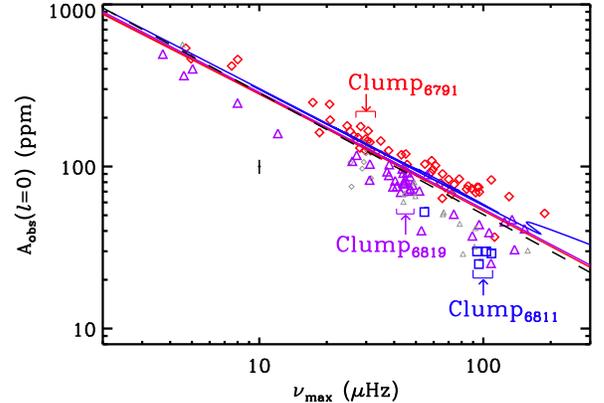}
\caption{Observed amplitude versus \numax~for stars
  in NGC~6791 (red diamonds), NGC~6819 (purple triangles), and NGC~6811
  (blue squares).  The binary stars are shown with small gray symbols.  The
  clump stars are marked. The dashed line shows a power law with slope
  $-0.75$.  Colored lines are the cluster isochrones (Figure~\ref{cmd})
  where amplitude and \numax~have been derived using Equation~\ref{scaling}
  with $s=0.75$ and
  \numax$=(M/M_\odot)/(L/L_\odot)(T_{\mathrm{eff}}/5777\mathrm{K})^{3.5}\,3100\,\mu$Hz.
  The black cross at (10,100) indicates a typical 1-$\sigma$ error bar. 
\label{ampnumax}} 
\end{figure} 
In Figure~\ref{ampnumax} we show the measured amplitude
as a function of \numax, where each set of symbols
present results of one cluster. 
We also mark the location of the clump of helium-core burning stars for each cluster, which
illustrates the large range in \numax~arising mainly from the difference in  
the stellar mass between the clusters.  

Guided by the fiducial dashed line, we see that stars within each
cluster roughly follow a power law with exponent $-0.75$, but with a clear
offset from one cluster to another by up to $\sim50$\%.  The more massive
the stars, the lower the oscillation amplitudes at a given \numax.  
This offset is not expected from the scaling relations for $A_\lambda$ and
\numax, as illustrated by the isochrones in Figure~\ref{ampnumax}.
Since the scaling
relation for \numax~is probably good to within a few percent
\citep{Stello09a,Belkacem11}, the observed offsets strongly
suggest that $(L/M)^sT_{\mathrm{eff}}^{-r}$ does not adequately predict the
amplitude for these stars.    
From a large sample of field red giants \citet{Huber10} noted that
the scatter in the amplitude at a given \numax~was larger than expected
from the uncertainties and that this indicated a spread in mass in their
sample.  However, a qualitative analysis was not attempted 
due to the relatively large uncertainties in the
fundamental stellar parameters.
Fortunately, with our cluster sample we can
directly fit the measured amplitudes
to their predictions derived from well-constrained stellar parameters. 

\subsection{Fitting the $L/M$ relation}
First, we fitted the observed amplitudes to the predicted 
amplitudes for NGC~6819.  For this purpose we derived the predicted
amplitude using the $L/M$ relation (Equation~\ref{scaling})
and adopting $\lambda=650\,$nm as the central wavelength of the
\kepler~bandpass, hence $A_\mathrm{obs,\odot}=3.49$ (peak scaled) \citep{Michel09}.  
The least-squares fit  
resulted in $s=0.76\pm0.01$ when adopting the empirical value of $r=2$,
which is the value of $r$ we will adopt in the following.  Using $r=1.5$ only has
the effect of increasing $s$ by about 0.03.
This result is compatible
with Paper I, which qualitatively
found the best match for $s$ to be slightly higher than 0.7. 
When repeated for NGC~6791, we found $s=0.87\pm0.01$. 
The small number of stars in NGC~6811 did not merit a fit on its own,
but the two other clusters already indicate inconsistent results.

Hence, we tried next fitting all three clusters
simultaneously.  Due to the correlation between $M$ and \teff~
(the hotter and younger clusters have more
massive stars; Figure~\ref{cmd}), we still kept
$r$ fixed. 
\begin{figure}
\includegraphics{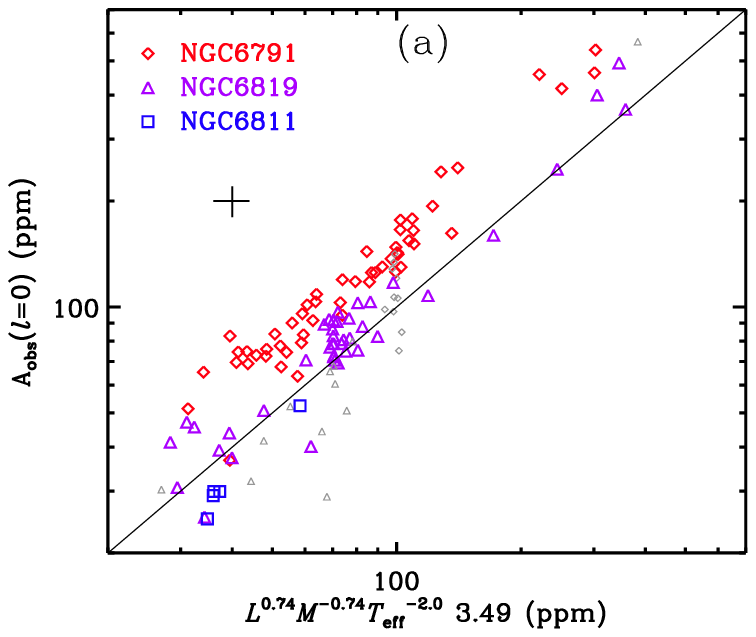}
\includegraphics{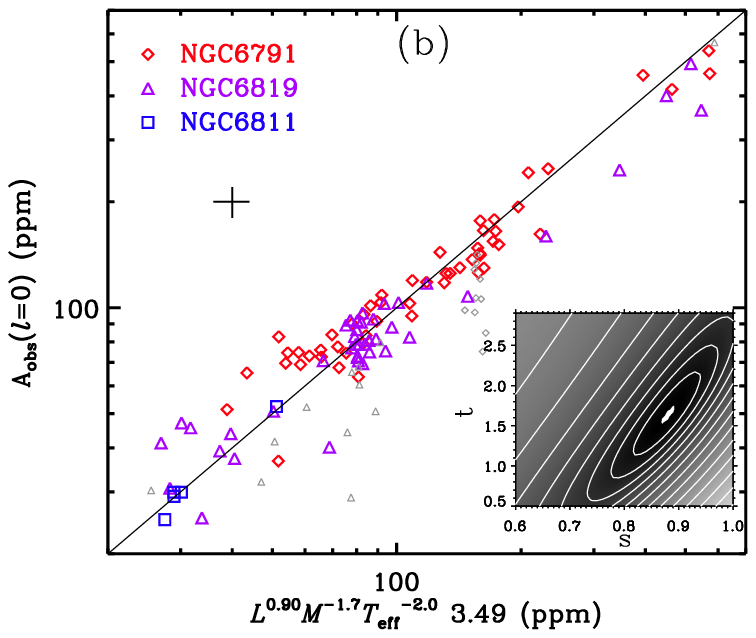}
\includegraphics{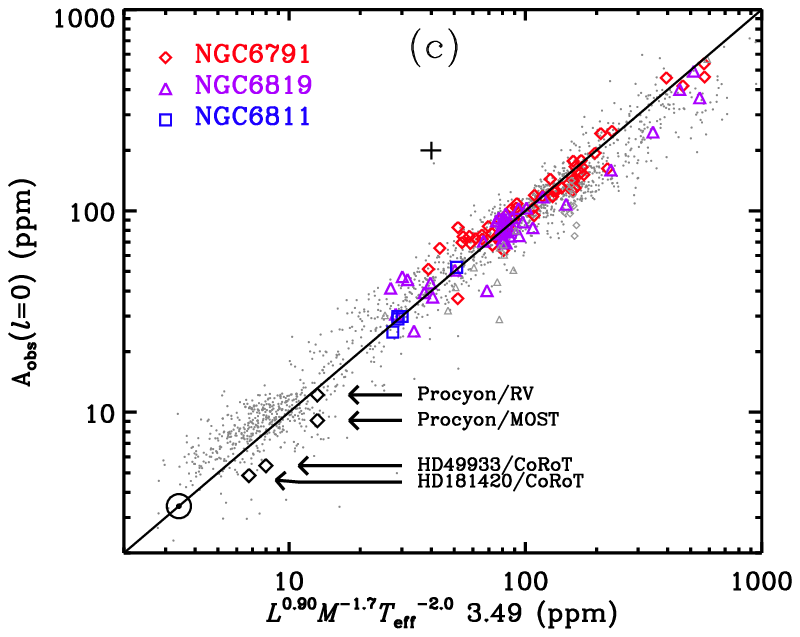}
\caption{(a) Observed versus predicted amplitude for the
  best fitting relation of the form $A_\lambda\propto (L/M)^sT_{\mathrm{eff}}^{-2}$.
  Symbols are the same as in Figure~\ref{ampnumax}.  Binaries, which are
  shown with small gray symbols, were not included in the fit.  
  (b) As panel (a) but fitting to $A_\lambda\propto L^sM^{-t}T_{\mathrm{eff}}^{-2}$.
  The inset shows the $\chi^2$ near its minimum. 
  (c) Illustration of how well the fit in panel (b) predicts
  amplitudes for other main-sequence, subgiant, and red giant stars
  (see text). 
\label{lmfit}} 
\end{figure} 
Figure~\ref{lmfit}(a) shows the result. 
The best fit resulted in $s=0.74\pm0.01$.
It is apparent that the clusters are offset from one
another, as expected from Figure~\ref{ampnumax}, but we also see that the
fit systematically underestimates the amplitude for the most luminous
stars.  
If $r$ was treated as a free
parameter we did obtain a better fit overall, but it still underestimated
the amplitudes of the stars in NGC~6791, and in particular the most
luminous stars in the sample, by 20--30\%. 
In summary, while the $(L/M)^s$ scaling provided acceptable results when
fitted to one cluster at a time (although giving different results for
$s$), our analysis has demonstrated 
that $(L/M)^s$ cannot explain the observations in
all clusters simultaneously.

\subsection{A new scaling relation for amplitudes}\label{new}
In the following, we therefore fitted the exponents on $L$ and $M$
independently,
hence $A_\lambda\propto L^sM^{-t}T_{\mathrm{eff}}^{-2}$.  The result, shown in 
Figure~\ref{lmfit}(b), is a much improved fit where all three clusters
fall on top of each other and follow the one-to-one relation.  The best fitting
parameters are $s=0.90\pm0.02$ and $t=1.7\pm0.1$ -- the same 
as we obtained from first converting $A_\mathrm{obs}$ to a bolometric
amplitude (Ballot et al.,
submitted) and then fitting to $A_{\mathrm{bol}}\propto L^sM^{-t}T_{\mathrm{eff}}^{-1}$.
For the stars with $A_\mathrm{obs}\gtrsim 80\,$ppm the scatter of
$A_\mathrm{obs}/(L^{0.90}M^{-1.7}T_{\mathrm{eff}}^{-2})$ is 14\%, in
perfect agreement with the quoted uncertainties on $A_\mathrm{obs}$, $L$,
$M$, and \teff.  
The increased scatter (22\%) towards lower luminosity stars is
potentially due to remaining issues of blending in the sample and/or an
increase in the uncertainties of the measured amplitudes for the faintest
stars.  The latter was, however, not reflected in the estimated uncertainties
reported by the pipelines showing only slightly increased uncertainties at
most.   
Again, under the adiabatic assumption ($r=1.5$) $s$ would slightly increase 
(to 0.95) as would $t$ (to 1.8). 

To investigate the robustness of our fit we did the following.
If we ignored the NGC~6811 stars in the fitting, the result and hence the
excellent alignment of all three clusters was very similar ($s$ and $t$
within $1\sigma$).  This is perhaps not surprising given the few stars in
our NGC~6811 sample.  Nevertheless, this result is reassuring 
since the amplitudes of NGC~6811 are then correctly predicted from a fit
based only on NGC~6791 and NGC~6819.
We further investigated the effect on the fit if we ignored all clump stars
to obtain an even more homogeneous sample, which showed practically no
change to the best fitting parameters. 
This indicates that any possible mass loss, which is expected to occur
predominantly near the tip of the red giant branch, has no effect on our
result.
A small systematic change of a few percent on $s$ and $t$ was, however,
observed by 
removing some of the most deviant stars at low amplitudes.
Finally, we repeated the fit on the sample of stars that were in common
between the SYD and CAN pipelines. The differences in $s$ and $t$ based on
these different pipelines were 2\% and
15\% in $s$ and $t$, respectively, the latter only just within $3\sigma$ of
the formal uncertainty.

We finally tested the new scaling relation suggested by
\citet{KjeldsenBedding11}, but found it to overestimate the amplitude for
the cluster stars similar to the result found by Huber et al. (submitted) and
Mosser et al. (submitted). 

\subsection{Main-sequence and subgiant stars}
Now, with an improved scaling relation for red giant stars, it is
interesting to see how well it applies to main-sequence and subgiant stars.
To investigate this we took amplitude measurements of the \kepler~field stars
presented by Huber et al. (submitted), the CoRoT F-type stars HD49933 and HD181420
from \citet{Michel08} (converted to $A_\mathrm{obs}(l$=0$)$), and Procyon
from \citet{Arentoft08} and \citet{Huber11}.  The amplitude measurement in
velocity of Procyon was converted to intensity 
using models by \citet{Houdek10}. We used our new scaling relation to
predict the amplitudes based on $L$, $M$, and \teff~from 
Huber et al. (submitted) (\kepler~sample), \citet{Bruntt09} (HD49933/181420), and
\citet{Bonanno07} (Procyon).  Given that the new relation is only based on
the cluster red giants, it is remarkable how well it agrees for
this broad range of stars (Figure~\ref{lmfit}c).  We note that the
uncertainty in the mass of 
the \kepler~($\sim 10$--20\%) and CoRoT ($\sim 5$--10\%) field stars
is significantly 
larger than for Procyon ($\sim 2$\%) and the cluster stars ($\sim
1$--2\%). 
While the values of $s$ and $t$ found in this Letter are slightly
different from, although still in agreement within the uncertainties, those
found by Huber et al. (submitted) for the \kepler~field stars, the qualitative
agreement across all stars is quite similar to that found by
Huber et al. (their Fig.~5).

\subsection{Unresolved binaries}
It is evident, particularly from Figure~\ref{lmfit}(b), that many of the
known and potential binaries (small gray diamonds and triangles) show
relatively low amplitudes.  For NGC~6791 we had no spectroscopic 
determination of binaries, but a significant fraction of its red clump
stars show 
lower than expected amplitudes and hence strong evidence for 'diluted' light
curves due to the presence of unresolved binary companions.  This shows a
new exciting way of applying asteroseismology to identify binary stars and
hence to probe the formation of these stars in clusters, which will
be investigated in detail in a forthcoming paper.

\section{Conclusions}
Our analysis of solar-like oscillations in 100 red giant stars in three
open clusters
revealed that previously adopted scaling relations based on the 
luminosity-to-mass ratio for predicting amplitudes are not adequate for red
giants. 
We found an empirical scaling relation by fitting the observed amplitudes
to a more general form than the previous $L/M$ relation. The result,
\begin{eqnarray}
A_\lambda &\propto& L^{0.90}/(M^{1.7}T_{\mathrm{eff}}^{2}) 
\end{eqnarray}
\noindent and
\begin{eqnarray}
A_\mathrm{bol} &\propto& L^{0.90}/(M^{1.7}T_{\mathrm{eff}}),
\end{eqnarray}
which showed considerable improvement for red giants, turned out to also
work remarkably well for main-sequence and subgiant stars. 

Interestingly, the lower than expected amplitudes of some red clump stars in
NGC~6791 and NGC~6819 revealed that they were likely unresolved 
binaries, many of which were not known previously.   
This method for identifying binaries could add interesting new insight to
the formation history of these clusters.

In this investigation we ignored any possible effect on amplitude from
metallicity differences \citep{Samadi07} of the three clusters, which have
values of [Fe/H]$_{\mathrm{NGC6791}}\simeq0.3$ and [Fe/H]$_{\mathrm{NGC6819}}\simeq0.1$
\citep[see~][and reference therein]{Basu11}, while it
is unknown for NGC~6811.  To improve on that will
require better determination of the cluster metallicities.  In addition, we
would need more clusters (with significantly different
stellar parameters) 
to allow the fitting of a rigorous empirical relation
including one more free parameter such as metallicity.

With more \kepler~data in the future, we expect to have cluster stars
covering a large range of \teff, which will include
turn-off stars at the end of the main sequence, allowing us to also fit the
exponent, $r$, of the \teff~dependence in the amplitude scaling relation.

\acknowledgments
Funding for this Discovery mission is provided by NASA's Science Mission
Directorate. 
We thank the entire {\it Kepler} team without whom this
investigation would not have been possible.
We acknowledge support from the: ARC, NWO, NSF, Polish Ministry and
Lend\"ulet program OTKA grants N-N203-405139, K83790, MB08C-81013.


\end{document}